\documentclass[twocolumn,showpacs,preprintnumbers,amsmath,amssymb]{revtex4}

\usepackage{graphicx}
\usepackage{dcolumn}% Align table columns on decimal point
\usepackage{bm}% bold math

\begin{document}

\preprint{APS/123-QED}

%\preprint{?? HEP/123-qed}

\title{Thermal Mechanism of Absolute Negative Conductivity\\
in Two-Dimensional Electron Systems 
} 
\author{ 
V.~Ryzhii }
\email{v-ryzhii@u-aizu.ac.jp}
%\cite{byline} 
%\author{XXXXX}
%\altaffiliation{Computer Solid State Physics Laboratory, University of Aizu,
%Aizu-Wakamatsu 965-8580, Japan} 
\affiliation{Computer Solid State Physics Laboratory, University of Aizu,
Aizu-Wakamatsu 965-8580, Japan}

%\author{ }
%\affiliation{

\date{\today}

\begin{abstract}
We calculate the dissipative dc conductivity
of a two-dimensional electron system in a magnetic field
for the situation when its effective temperature exceeds
the temperature of the acoustic phonon system.
We demonstrate that at sufficiently large difference between
the  electron and lattice temperatures
(sufficiently strong microwave radiation)
the dissipative dc conductivity can become negative.
As an illustration, the case of
 the electron heating by microwave radiation is considered.
\end{abstract}



\maketitle

\section{Introduction}

Recent experimental studies of new types of magnetoresistance
oscillations in a two-dimensional  electron system (2DES)
in a magnetic field~\cite{1,2,3,4,5},
particularly  the observation of
vanishing electrical resistance 
caused  by microwave radiation, have stimulated 
extensive efforts to clarify the  origin
of the uncovered phenomena.
The occurrence of the effect of the so-called zero-resistance
states  
is closely associated with
the effect of  absolute negative conductivity (ANC)
when the dissipative dc conductivity $\sigma$ is negative~\cite{6,7,8}.
Mechanisms of ANC in a 2DES subjected to a magnetic field
under nonequilibrium conditions, in particular
under 
irradiation with microwaves, have been studied theoretically
since the late 60s~\cite{9,10,11,12,13,14,15,16,17}.
One can speculate that the photon-assisted impurity scattering~\cite{11,12,13}
is the main mechanism of ANC causing the zero-resistance states.
%%%%%
Notwithstanding  this, a revision  of alternate possible mechanisms of ANC
may be of interest as well.

In this paper, we show that ANC can occur owing to the electron-phonon
scattering (acoustical)
if a 2DES is not in equilibrium with the phonon system.
The deviation of the 2DES and the acoustic phonon systems from equilibrium
can be caused, in particular, by the electron heating 
due to the absorption of microwaves or infrared radiation
accompanied by  the 
transitions between the Landau levels (LL's) or 
the intersubband transitions.
This is consistent with the concept considered previously~\cite{9,10}.
Here we derive the  expression for
 the dc dissipative component
of the 2DES conductivity $\sigma$ as a function of the phonon (lattice) and
2DES (electron) temperatures, $T$ and $T_e$, respectively,
 considering
the electron interaction with acoustic phonons. After that, we calculate
 $T_e$ for the case of the electron heating by microwave
radiation due to the absorption
 associated with the photon-assisted electron 
scattering on the phonons of the same type. 
We show that at sufficiently high electron temperatures
(sufficiently strong radiation), the dissipative dc conductivity
associated with the electron-phonon interaction
can become negative, i.e., the electron heating can lead to the effect of ANC.
The thermal mechanism of ANC under consideration can markedly influence
the transport properties of a 2DES affected by microwave or infrared
radiation.

\section{Conductivity}

The dissipative dc current $j_D(E)$
in a
2DES in the case of the electron-phonon interactions
 can be calculated
from the following quite general expression:
$$
j_D(E) = \frac{e}{\hbar} \sum_{N,N^{\prime} }
f_N(1 - f_{N^{\prime}})
$$
$$
\times\int d^3{\bf q}\,
q_y|V_{{\bf q}}|^2|Q_{N,N^{\prime}}(L^2q^2_{\perp}/2)|^2
$$
$$
\times\{
{\cal N}_q
\delta[(N - N^{\prime})\hbar\Omega_c + \hbar\omega_q + eEL^2q_y]
$$
\begin{equation}\label{eq1}
+ ({\cal N}_q + 1)
\delta[(N - N^{\prime})\hbar\Omega_c - \hbar\omega_q + eEL^2q_y] 
\}.
\end{equation}
Here,
$L = (c\hbar/eH)^{1/2}$
is the quantum Larmor radius, 
$\Omega_c = eH/mc$ is the cyclotron frequency,
$f_N$ and  ${\cal N}_q$,
are the electron and phonon
distribution functions, respectively, 
$E$ is the net in-plane electric field,
$H$ is the magnetic field,
$N = 0.1,2,...$ is the  LL index,
 ${\bf q} = (q_x, q_y, q_z)$ (the directions $x$ and $y$ are 
in the 2DES plane), ${\bf q}_{\perp} = (q_x, q_y)$,
 and $\omega_q = sq$
are 
the phonon wave vector, its in-plane component,  and the phonon frequency, 
respectively,
 $e = |e|$ is the electron charge, $\hbar$ is the Planck constant,
$c$ and $s$ are the velocities
of light and sound, respectively,
$\delta (\omega)$ is  the
 form-factor of LL's which at a small  broadening 
$\Gamma$ can be assumed to be the Dirac delta function,
$V_{{\bf q}}$ is the matrix element of the electron-phonon interaction, 
$|Q_{N,N^{\prime}}(L^2q^2_{\perp}/2)|^2  = 
|P_N^{N^{\prime} - N}(L^2q^2_{\perp}/2)|^2 \exp (- L^2q^2_{\perp}/2))$ 
is determined by the overlap of the electron wave functions, and
$|P_N^{N^{\prime} - N}(L^2q^2_{\perp}/2)|^2$
is proportional to a Laguerre  polynomial.
Considering the acoustic piezoelectric scattering, we set
$|V_{{\bf q}}|^2$$ \propto q^{-1}\exp(- l^2q_z^2/2)$,
where
 $l$ is
the electron localization length in the $z$-direction perpendicular to
the 2DES plane. Usually  $l \ll L$.  
We shall assume that the electron and phonon distributions
are characterized by the respective temperatures $T_e$ and $T$:
$f_N = [\exp(N\hbar\Omega_c - \zeta)/T_e]^{-1},\,   
{\cal N}_q = [\exp(\hbar\omega_q/T)  - 1]^{-1}$,
where $\zeta$ is the Fermi energy reckoned from the lowest LL.

Assuming that 
$eEL \ll   \hbar \Omega, \hbar\omega_q$, 
one can expand  the expression in the right-hand side 
of Eq.~(1) in powers of $(eEL^2/\hbar s)$ and present 
the dissipative conductivity $\sigma = j_D/E$
in the following form:
$$
\sigma = \biggl(\frac{e^2L^2}{\hbar^3s^2}\biggr) 
\sum_{N,\Lambda}
\int d^3{\bf q}\,
q_y^2|V_{{\bf q}}|^2
$$
$$
\times|Q_{N,N+\Lambda}(L^2q^2_{\perp}/2)|^2
\delta^{\prime}(q  - q^{\Lambda})
$$
\begin{equation}\label{eq2}
\times\{[f_N(1 - f_{N+\Lambda}){\cal N}_q - f_{N+\Lambda}(1 - f_N)({\cal N}_q + 1)],
\end{equation}
where $\delta^{\prime}(q) = d\delta(q)/dq$ ,
$q^{(\Lambda)} = \Lambda\Omega_c/s$, and $\Lambda > 0$.
After replacing  the integration over $d^3{\bf q}$ by the integration
over $dq\,dq_{\perp}d\theta$, where $\sin \theta = q_y/q_{\perp}$,
Eq.~(3) 
can be reduced to 
$$
\sigma \propto 
\sum_{N\Lambda}
\int_0^{\infty}
\frac{dqdq_{\perp}q_{\perp}^3}
{\sqrt{q^2 - q_{\perp}^2}}\exp\biggl[- \frac{l^2(q^2 - q_{\perp}^2)}{2}\biggr]
$$
$$
\times|Q_{N,N+\Lambda}(L^2q^2_{\perp}/2)|^2\delta^{\prime}(q  - q^{(\Lambda)})
$$
\begin{equation}\label{eq3}
\times
[f_N(1 - f_{N+\Lambda}){\cal N}_q - f_{N+\Lambda}(1 - f_N)({\cal N}_q + 1)],
\end{equation}
Substituting 
the explicit formulas for the electron
and phonon distributions 
functions  into Eq.~(3), 
one can  thereafter   present it as

$$
\sigma  \propto 
\sum_{N,\Lambda}f_N(1 - f_{N+\Lambda})
\int_0^{\infty}
dq 
\exp\biggl(- \frac{l^2q^2}{2}\biggr)
G_{N}^{\Lambda}(q)
$$
\begin{equation}\label{eq4}
\times
\frac{1 - \exp[\hbar (sq/T - \Lambda\Omega_c/T_e)]}
{\exp(\hbar s q/T) - 1}
\delta^{\prime}(q  - q^{(\Lambda)}),
\end{equation}
where 
$$
G_N^{(\Lambda)}(q) = \int_0^{Lq} dtt^3
\frac{\displaystyle\exp[(l^2 - L^2)t^2/2L^2]}
{\sqrt{L^2q^2 - t^2}}
|P_N^{\Lambda}(t^2/2)|^2.
$$
Assuming that $T, T_e \ll \hbar\Omega_c$, and 
neglecting
the terms containing higher 
powers of $\exp(- \hbar\Omega_c/kT)$, 
we arrive at

$$
\sigma  \propto 
\int_0^{\infty}
dq\exp(- l^2q^2/2)
G(q)
$$
\begin{equation}\label{eq5}
\times[\exp(-\hbar sq/T) - \exp(- \hbar\Omega_c/T_e )]
\delta^{\prime}(q  - q^{(1)}),
\end{equation}
where $G(q) = f_{N_m}(1 -  f_{N_m + 1})G_{N_m}^{(1)}(q)$
and
$N_m$ in the index of the LL
immediately below the Fermi level.
The integration in Eq.~(5) gives
\begin{equation}\label{eq6}
\sigma  \propto \exp(- \hbar\Omega_c/T)\frac{d[
\exp(- l^2q^2/2)G(q)F(q)]}{dq}\biggr|_{q = q^{(1)}}.
\end{equation}
Here 
$$
F(q) = \frac{\exp(- \hbar\Omega_c/T_e) - 
\exp(- \hbar sq/T)}{\exp(- \hbar\Omega_c/T)},
$$

At $Lq \gg 1$, one obtains
$G_N^{(\Lambda)}(q) \simeq (1 -l^2/L^2)^{3/2}\overline{G}_N^{(\Lambda)}/Lq
\simeq \overline{G}_N^{(\Lambda)}/Lq$.
If $1/\sqrt{N} < Lq < 1$, one obtains
$G_N^{(\Lambda)}(q)\simeq \tilde{G}_N^{(\Lambda)}(Lq)^2$.
Here $\overline{G}_N^{(\Lambda)}$ and $\tilde{G}_N^{(\Lambda)}$
are coefficients. 
Taking into account that $Lq^{(1)} = L\Omega_c/s \gg 1$ and, hence,
 $G(q) \simeq   \overline{G}_N^{(\Lambda)}/Lq$ at $q \simeq q^{(1)}$,
from  Eq.~(6) we obtain
$$
\frac{\sigma}{\sigma_0} =1 + \biggl(\frac{T}{\hbar\Omega_c}\biggr)
\biggl(1 + \frac{l^2\Omega_c^2}{s^2}\biggr)
$$
\begin{equation}\label{eq7}
\times\biggl[1 - \exp\biggl(\frac{\hbar\Omega_c}{T} - \frac{\hbar\Omega_c}{T_e}\biggr)
\biggr].
\end{equation}
Here  $\sigma_0$  is the conductivity at $T_e = T$ given
by the following formula~\cite{12}:
\begin{equation}\label{eq8}
\sigma_0 \propto 
\biggl(\frac{\hbar s}{TL}\biggr)\biggl(\frac{s}{L\Omega_c}\biggr)
\exp\biggl(- \frac{\hbar\Omega_c}{T}\biggr)
\exp\biggl(- \frac{l^2\Omega_c^2}{2s^2}\biggr).
\end{equation}
At a small electron localization length
$l$, the last exponential factor in the right-hand side
can be approximately  replaced by unity.

Using Eq.~(7), one can show that $\sigma$ becomes negative when

\begin{equation}\label{eq9}
T_e  > \frac{T}{1 - \theta} = T_{t},
\end{equation}
where
$\theta \simeq (T/\hbar\Omega_c)\ln(\hbar\Omega_c/T)$.
The origin of ANC in such a situation is due to the following.
When the electron temperature exceeds the lattice temperature,
the balance between  the processes of the phonon absorption
accompanied by  electron transitions to higher LL's
and  the processes of the phonon emission accompanied by
electron transitions to lower LL's
is violated, and the processes of the latter type dominate.
The processes associated with the emission of a phonon
with energy less than the LL separation provide
a negative contribution to the dissipative conductivity, whereas
those with the energy exceeding this separation yield
 a positive contribution.
Since  the probability
of the emission of a phonon with a lower energy
is larger, the net contribution of the electron scattering processes
involving the emission of phonons can be negative  if
the electron temperature  is sufficiently large compared to
the lattice temperature.

\section{Effective temperature at microwave heating}

Let us consider the situation when the 2DES under consideration
is heated by incoming microwave radiation.
In this case, 
the electron distribution over LL's is determined by
the balance between the electron relaxation
associated with the emission of acoustic phonons and the processes
of the electron heating caused by the photon-assisted
absorption/emission of such phonons.
The balance equation can be presented in the following form:

$$
\hbar s\sum_{N}\sum_{\Lambda}
f_N(1 - f_{N+\Lambda})
\int_0^{\infty}
dq\,q 
\exp(- l^2q^22)
K_{N}^{(\Lambda)}(q)
$$
$$
\times[\exp( - \Lambda\hbar\Omega_c/T_e) ({\cal N}_q + 1) - {\cal N}_q]
\delta(q - q^{(\Lambda)})
$$
$$
= \hbar\Omega{\cal J}_{\Omega}\sum_{N}\sum_{\Lambda}
f_N(1 - f_{N+\Lambda})
\int_0^{\infty}
dq 
\exp(- l^2q^22)
G_{N}^{\Lambda}(q)
$$
\begin{equation}\label{eq10}
\times[{\cal N}_q\delta(q + q^{(\Lambda)}_{\Omega})
+ ({\cal N}_q + 1)\delta(q - q^{(\Lambda)}_{\Omega})]
\end{equation}
Here $q^{(\Lambda)}_{\Omega} \ (\Omega - \Lambda\Omega_c)/s$,
${\cal J}_{\Omega} = ({\cal E}_{\Omega}/\tilde{{\cal E}}_{\Omega})^2$,
and ${\cal E}_{\Omega}$ and
$\tilde{{\cal E}_{\Omega}}$  are the microwave electric field amplitude
and its characteristic value, respectively. 
The right-hand side of Eq.~(10) is valid at sufficiently low microwave
powers when the multi-photon processes (both real and virtual)
are insignificant.  
At relatively low
microwave powers when
${\cal E}_{\Omega} \ll \tilde{{\cal E}_{\Omega}}$, disregarding the polarization
anisotropy, one can use the following formula~\cite{18,19}: 
$\tilde{{\cal E}}_{\Omega} = \overline{{\cal E}}_{\Omega}|\Omega_c^2 - \Omega^2|L/
[e\Omega\sqrt{\Omega_c^2 +  \Omega^2}]$,
where $ \overline{{\cal E}}_{\Omega} = \sqrt{2}m\Omega^2L/e$.
At $Lq \gg 1$, the function $K_{N}^{\Lambda}(q) \simeq 
\overline{K}_{N}^{\Lambda}/Lq$  with the coefficients $\overline{K}_{N}^{\Lambda}$
slightly different from $\overline{G}_{N}^{\Lambda}$.
In the vicinity of the cyclotron resonance $\Omega = \Omega_c$,
the quantity $\tilde{{\cal E}_{\Omega}}$ is limited by the LL broadening
if ${\cal E}_{\Omega}/ \overline{{\cal E}_{\Omega}} < \Gamma/\Omega_c$.
In this limit one can put $\tilde{{\cal E}_{\Omega}} 
\simeq \sqrt{2}m\Omega\Gamma/e \overline{{\cal E}}_{\Omega}\Gamma/\Omega_c$.
In the case  ${\cal E}_{\Omega}/ \overline{{\cal E}_{\Omega}} > \Gamma/\Omega_c$,
when the multi-photon processes  (both real and virtual) are
essential,
the situation becomes more complex~\cite{18,19}.

Near the $\Lambda$th resonance 
$|\Omega - \Lambda\Omega_c|\lesssim s/L,\,T/\hbar \ll \Omega_c$,
Eq.~(10) yields
\begin{equation}\label{eq11}
\frac{T_e - T}{T} \simeq 
{\cal J}_{\Omega}\frac{L^2(\Omega - \Lambda\Omega_c)^2}{s^2}
\biggl(\frac{TL}{\hbar s}\biggr)\exp\biggl(\frac{\hbar\Omega}{T}\biggr)
\Lambda\tilde{\Phi}_{\Lambda}.
\end{equation}
Here
$$
\tilde{\Phi}^{(\Lambda)} = \tilde{M}^{(\Lambda)}\biggl(\frac{TL}{\hbar s}\biggr)
\exp\biggl(\frac{l^2\Omega_c^2 }{2s^2}\biggr)
\simeq \tilde{M}^{(\Lambda)}
$$
and
$$
 \tilde{M}^{(\Lambda)} = \frac{\displaystyle\sum_{N = N_m -\Lambda+1}^{N_m}\tilde{G}_{N}^{(\Lambda)}f_{N}(1 -  f_{N + 1})
}
{\displaystyle\overline{K}_{N_m}^{(1)}f_{N_m}(1 -  f_{N_m + 1})} \propto \Lambda.
$$
One can see that the right-hand side of Eq.~(11) tends to zero
when the resonance detuning $\Omega - \Lambda\Omega_c$ approaches  zero.
This occurs even at the cyclotron resonance $\Lambda = 1$ since an increase
in ${\cal J}_{\Omega}$
at fixed microwave intensity  
is limited by the LL broadening:
${\cal J}_{\Omega} \propto [(\Omega - \Omega_c)^2 + \Gamma^2]^{-1}$.
Therefore, near the cyclotron resonance 
\begin{equation}\label{eq12}
\frac{T_e - T}{T} \propto
\frac{(\Omega - \Omega_c)^2}{[(\Omega - \Omega_c)^2 + \Gamma^2]}.
\end{equation}
Thus, at the resonances, the electron heating 
associated with the photon-assisted acoustic scattering 
does not take place. In this case, the electron heating
can occur due to other photon-assisted events. 
%%%%%%%
It is instructive that  the dependence of the electron temperature on
the resonance detuning $(\Omega - \Lambda\Omega_c)$ is virtually
an even function of the latter if $|\Omega - \Lambda\Omega_c|\lesssim s/L,\,T/\hbar$.
When the resonance detuning is sufficiently large
$s/L,\,T/\hbar \lesssim |\Omega - \Lambda\Omega_c| < \Omega_c$, 
the electron temperature becomes a markedly asymmetric
function of $(\Omega - \Lambda\Omega_c)$, particularly at very low lattice
temperatures.
Indeed, in such a range of  detuning, 
from Eq.~(10) we obtain the following relationships
for the electron temperature:
%%%%%%%%%%%%%%%%
$$
\exp\biggl(\frac{\hbar\Omega_c}{T} - \frac{\hbar\Omega_c}{T_e}\biggr)
= 1 
$$
\begin{equation}\label{eq13}
+ {\cal J}_{\Omega}
\frac{\Omega}{|\Omega - \Lambda\Omega_c|}
\exp\biggl[\frac{\hbar(\Omega - \Lambda\Omega_c) }{T}\biggr]
\exp\biggl(\frac{\hbar\Omega_c }{T}\biggr)\overline{\Phi}^{(\Lambda)}
\end{equation}
at $\Omega - \Lambda\Omega_c < 0$, and 
$$
\exp\biggl(\frac{\hbar\Omega_c}{T} - \frac{\hbar\Omega_c}{T_e}\biggr)
= 1 
$$
\begin{equation}\label{eq14}
+ {\cal J}_{\Omega}
\frac{\Omega}{|\Omega - \Lambda\Omega_c|}
\exp\biggl(\frac{\hbar\Omega_c }{T}\biggr)\overline{\Phi}^{(\Lambda)}
\end{equation}
at $\Omega - \Lambda\Omega_c > 0$,
where $\overline{\Phi}^{(\Lambda)}$ is similar to $\tilde{\Phi}^{(\Lambda)}$.

\section{Spectral dependences}

Using Eqs.~(7) and (11)
and assuming for simplicity that 
$l\Omega_c/s < 1$, $TL/\hbar s \simeq 1$, 
$\tilde{\Phi}^{(\Lambda)} \simeq \Lambda$, and
$\overline{\Phi}^{(\Lambda)} \simeq \Lambda$, we obtain for the 
dissipative conductivity 
near the $\Lambda$th resonance

\begin{equation}\label{eq15}
\frac{\sigma_0 - \sigma}{\sigma_0}\simeq  {\cal J}_{\Omega}
\frac{\Lambda^2L^2(\Omega - \Lambda\Omega_c)^2}{s^2}
\exp\biggl(\frac{\hbar\Omega_c }{T}\biggr).
\end{equation}
Using Eqs.~(7) and  (13),
at the microwave frequencies well below the 
cyclotron resonance $\Omega < \Omega_c$, we arrive at the following formula:
\begin{equation}\label{eq16}
\frac{\sigma_0 - \sigma}{\sigma_0}\simeq {\cal J}_{\Omega}
\frac{\Omega}{|\Omega - \Omega_c|}
\exp\biggl[\frac{\hbar(\Omega - \Omega_c) }{T}\biggr]
\exp\biggl(\frac{\hbar\Omega_c }{T}\biggr).
\end{equation}
At higher microwave frequencies $\Omega_c \lesssim \Omega < 2\Omega_c$, Eqs.~(7), (13), and (14)
yield
\begin{equation}\label{eq17}
\frac{\sigma_0 - \sigma}{\sigma_0} \simeq  {\cal J}_{\Omega}\frac{\Omega}
{|\Omega - \Omega_c|}
\exp\biggl(\frac{\hbar\Omega_c }{T}\biggr).
\end{equation}
The obtained dependences of $\sigma$ versus $\Omega/\Omega_c$
exhibit deep minima on the either side of the cyclotron resonances
with $\sigma < 0$ if the intensity of microwave radiation
is sufficiently large. The positions of these minima
are $\Omega^{(\mp)} = \Omega_c \mp \delta_m$ with $\delta_m \simeq s/L$.
Similar minima, while less pronounced,  occur at  higher resonances.
Comparing Eqs.~(16) and (17), we obtain the following
estimate: $|\sigma_0 - \sigma^{(+)}|/|\sigma_0 - \sigma^{(-)}| 
\simeq \exp(\hbar s/LT)$.
At sufficiently low lattice temperatures, 
this ratio is large because the contribution
to the electron heating from the photon-assisted absorption of acoustic phonons
is rather small. 
This implies that the  dissipative dc conductivity 
at the microwave frequencies somewhat below the resonances (given by Eq.~(7))
is close to its ``dark'' value $\sigma_0$.
The dependences of the normalized
 dissipative dc conductivity $\sigma/\sigma_0$ on $\Omega/\Omega_c$
calculated numerically using Eqs.~(7) and (10) for different values
of the parameter $b = \hbar s/LT$ 
are shown in Fig.~1  For the  numerical calculations we 
assumed that $H = 2$~kG,  $l/L =0.1$,  $\gamma =\Gamma/\Omega_c = 0.1$
and $S = ({\cal E}_{\Omega}/ \overline{{\cal E}_{\Omega}})^2 = 10^{-3}$.
Figure~2 shows the normalized
 dissipative  conductivity $\sigma/\sigma_0$ as a function
of the parameter $b \propto T^{-1}$ at different values of the detuning
$\Omega - \Omega_c$.
One can see from Figs.~1 and 2
that at sufficiently large $b$, i.e., at sufficiently
low lattice temperatures, 
the ratio $\sigma/\sigma_0$ as a function of $\Omega$ and $\Omega_c$ 
exhibits pronounced minima
at which it can be negative.

\begin{figure}
\begin{center}
\includegraphics[width=7.5cm]{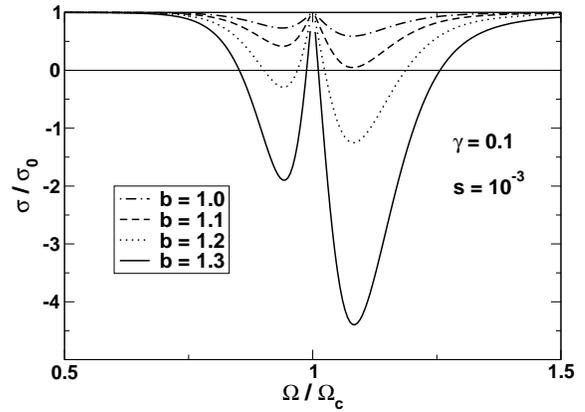}
\end{center}
\label{fig1}
\caption{
Normalized dissipative conductivity vs microwave 
frequency
for different parameter $b$
(inverse lattice temperature).
}
\end{figure} 
\begin{figure}
\begin{center}
\includegraphics[width=7.5cm]{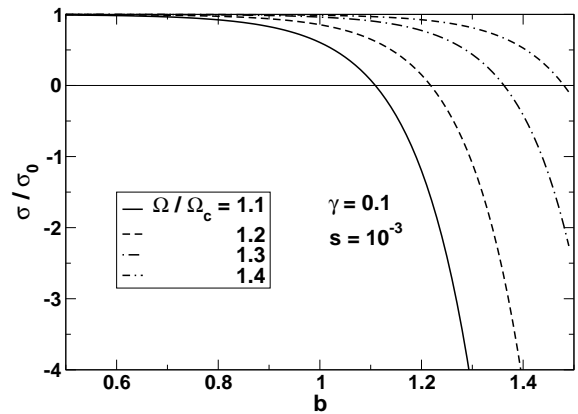}
\end{center}
\label{fig2}
\caption{Normalized dissipative conductivity vs parameter $b$ 
for different resonance detuning.
}
\end{figure}

Taking into account Eq.~(17) and setting $\Omega - \Omega_c = s/L$,
one can
obtain the following condition for ANC associated with
the mechanism under consideration:
\begin{equation}\label{eq18}
{\cal J}_{\Omega} > \biggl(\frac{s}{L\Omega}\biggr)
\exp\biggl(- \frac{\hbar\Omega_c }{T}\biggr).
\end{equation}
The right-hand side of inequality~(18) comprises two small parameters:
$s/L\Omega$  and $\exp(- \hbar\Omega_c/T)$. For   a  2DES in
a GaAs/AlGaAs heterostructure,
assuming $s = 3\times10^5$~cm/s, 
$H = 2$~kG, and $T = 1$~K, the product of these parameters is estimated as
$10^{-3}$. 
Naturally, when the dark conductivity is determined by another
mechanism than the acoustic phonon scattering,
the occurrence  of ANC requires a larger microwave power.

\section{Discussion}

The electron interaction with interface acoustic phonons
can also contribute to the mechanism of ANC under consideration.
However, the dissipative  conductivity associated with
the electron scattering on 2D acoustic phonons is smaller than that
associated with the scattering on 3D phonons by a large factor
$\exp[(L^2 - l^2)\Omega_c^2/2s^2]$~\cite{17}. Due to this, the variation
of the dissipative conductivity 
caused by the electron heating is rather small. On the contrary, the contribution
of the photon-assisted electron interaction with 2D phonons
to the absorption of microwave radiation and, therefore, to the electron
heating can be pronounced. The latter (extra) heating mechanism can result
in an additional deepening of the conductivity  minima.

Thus, the electron
interaction with acoustic phonons
can lead to ANC in 
a 2DES irradiated with microwaves. As shown in Refs.~\cite{16,17} and
here,  two  ``acoustic'' mechanisms of ANC are possible:
(1) the dynamic mechanism associated with the direct contribution
of the photon-assisted scattering processes to the dissipative
conductivity and (2) the thermal mechanism at which the microwave radiation
increases the electron temperature that, in turn,  affects the dissipative
conductivity. Comparing the dynamic and thermal variations
of the dissipative  conductivity caused by irradiation, 
one can determine that their ratio
does not include any parameter significantly different from unity.
Hence, both the mechanisms in question can  be simultaneously essential. 
However, these mechanisms lead to substantially different
spectral behavior of the dissipative
conductivity. 

At the electron  heating
associated with the intersubband transitions
stimulated by infrared radiation, the  dissipative
conductivity can also become negative,
exhibiting, however, a rather smooth (nonresonant)
dependence on the magnetic field. 
The mechanism of ANC considered above is due to a difference
between the the electron and phonon
temperatures. Therefore, the dissipative conductivity
in a 2DES subjected to a magnetic field can be negative
not only at $T_e > T $ but at $T_e < T$ as well (see Ref.~\cite{10}).

\section{Summary}

We demonstrated
a possibility of the thermal mechanism of negative
dissipative dc conductivity
in a 2DES subjected to  a magnetic field
associated with the interaction of electrons
with acoustic phonons
at a difference between  the electron and phonon
temperatures. 
As an example, 
the electron heating by microwave radiation is considered.
It was shown that in this case, both dynamic and thermal 
mechanisms can be efficient to result in  ANC
at sufficiently strong microwave radiation in certain ranges
of its frequency.

\section*{Acknowledgments}

The author is grateful to V.~A.~Volkov and V.~V.~Vyurkov for useful discussions
and A.~Satou for assistance.

\end{document}